\documentclass[twoside,12pt]{article}
\usepackage{epsf}
\protect

\begin{document}

\begin{titlepage}
\noindent
\renewcommand{\thefootnote}{\fnsymbol{footnote}}
\parbox{1.85cm}{\epsfxsize=1.85cm \epsfbox{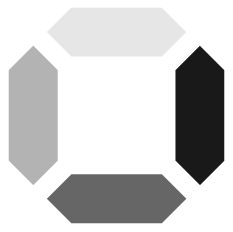}}\hfill%
\begin{minipage}{10cm}
\rightline{cond-mat/0001345}
\rightline{University of Essen}%
\rightline{January 24, 2000}%
\end{minipage}
\vfill 
\centerline{\sffamily\bfseries\Large Polymerized  Membranes, a Review}
\vfill
\centerline{\bf\large Kay J\"org Wiese%
\footnote{Email: wiese@next23.theo-phys.uni-essen.de}}
\smallskip
\centerline{\small Fachbereich Physik, Universit\"at GH Essen,  45117 Essen,
Germany}
\smallskip\smallskip

\vfill
\vspace{-5mm}
\begin{abstract}
Membranes are of great technological and biological as well as theoretical
interest. Two main classes of membranes can be distinguished:
Fluid membranes and polymerized, tethered membranes.  Here, we 
 review progress in the theoretical understanding of polymerized
 membranes, i.e.\ membranes with a fixed internal connectivity. 
We start by collecting basic physical properties,
clarifying the role of bending rigidity
 and disorder,  theoretically and experimentally as well as 
 numerically. 
We then give a thorough introduction into the theory 
of self-avoiding membranes, or more generally non-local field
theories with $\delta$-like interactions. 
Based on a proof of perturbative renormalizability
for non-local field-theories, renormalization group
calculations can be  performed up to 2-loop order, which in 3 dimensions 
predict a crumpled phase with fractal dimension of about 2.4; 
this phase is however
seemingly unstable towards the inclusion of bending rigidity. 
The tricritical behavior of membranes is discussed and shown 
to be quite different from that of polymers. 
Dynamical properties are studied in the same frame-work.
Exact scaling relations, suggested but not demonstrated long time ago 
by De Gennes for polymers, are established.
 Along the same lines, disorder can be 
included leading to interesting applications.
We also construct a generalization of the $O(N)$-model, which 
in the limit $N\to0$ reduces to self-avoiding membranes in analogy with the
$O(N)$-model, which in the limit $N\to0$ reduces to self-avoiding polymers.
Since perturbation theory is at the basis of the above approach,
one has to ensure that the perturbation
expansion is not divergent or  at least Borel-summable. Using a suitable
reformulation of the problem, we obtain the instanton governing
the large-order behavior. 
This suggest that the perturbation expansion is indeed
Borel-summable and the presented approach meaningful.  
Some technical details are relegated to the appendices.
A final collection of various topics  may also serve as exercises.

\end{abstract}
\vspace{-5mm}
\vfill

\centerline{\em Review to be published in volume 19 of Domb Lebowitz,} 
\centerline{\em Phase Transitions and Critical Phenomena.}
\smallskip
\centerline{\em Habilitation thesis, Fakult\"at f\"ur Physik, Universit\"at Essen, Germany}
\centerline{\em submitted May 21, 1999; accepted October 20, 1999.}
\vfill

\end{titlepage}
\renewcommand{\thefootnote}{\fnsymbol{footnote}}

\centerline{\bfseries \sffamily\large Table of Contents}
\setcounter{tocdepth}{4}
\contentsline {section}{\numberline {1}Introduction and outline}{5}
\contentsline {section}{\numberline {2}Basic properties of membranes}{9}
\contentsline {subsection}{\numberline {2.1}Fluid membranes}{9}
\contentsline {subsection}{\numberline {2.2}Tethered (polymerized) membranes}{10}
\contentsline {subsection}{\numberline {2.3}Crumpling transition, the role of bending rigidity, and some approximations}{12}
\contentsline {subsection}{\numberline {2.4}Stability of the flat phase}{14}
\contentsline {subsection}{\numberline {2.5}Experiments on tethered membranes}{19}
\contentsline {subsection}{\numberline {2.6}Numerical simulations of self-avoiding membranes}{21}
\contentsline {subsection}{\numberline {2.7}Membranes with intrinsic disorder}{23}
\contentsline {section}{\numberline {3}Field theoretic treatment of tethered membranes}{24}
\contentsline {subsection}{\numberline {3.1}Definition of the model, observables, and perturbation expansion}{24}
\contentsline {subsection}{\numberline {3.2}Locality of divergences}{27}
\contentsline {subsection}{\numberline {3.3}More about perturbation theory}{28}
\contentsline {subsection}{\numberline {3.4}Operator product expansion (OPE), a pedagogical example}{29}
\contentsline {subsection}{\numberline {3.5}Multilocal operator product expansion (MOPE)}{33}
\contentsline {subsection}{\numberline {3.6}Evaluation of the MOPE-coefficients}{35}
\contentsline {subsection}{\numberline {3.7}Strategy of renormalization}{39}
\contentsline {subsection}{\numberline {3.8}Renormalization at 1-loop order}{39}
\contentsline {subsection}{\numberline {3.9}Non-renormalization of long-range interactions}{44}
\contentsline {section}{\numberline {4}Some useful tools and relation to polymer theory}{45}
\contentsline {subsection}{\numberline {4.1}Equation of motion and redundant operators}{45}
\contentsline {subsection}{\numberline {4.2}Analytic continuation of the measure}{48}
\contentsline {subsection}{\numberline {4.3}IR-regulator, conformal mapping, extraction of the residue, and its universality}{50}
\contentsline {subsection}{\numberline {4.4}Factorization for $D=1$, the Laplace De Gennes transformation}{52}
\contentsline {section}{\numberline {5}Proof of perturbative renormalizability}{56}
\contentsline {subsection}{\numberline {5.1}Introduction}{56}
\contentsline {subsection}{\numberline {5.2}Proof}{57}
\contentsline {subsection}{\numberline {5.3}Some examples}{70}
\contentsline {section}{\numberline {6}Calculations at 2-loop order}{74}
\contentsline {subsection}{\numberline {6.1}The 2-loop counter-terms in the MS scheme}{74}
\contentsline {subsection}{\numberline {6.2}Leading divergences and constraint from renormalizability}{75}
\contentsline {subsection}{\numberline {6.3}Absence of double poles in the 2-loop diagrams}{77}
\contentsline {subsection}{\numberline {6.4}Evaluation of the 2-loop diagrams}{78}
\contentsline {subsection}{\numberline {6.5}RG-functions at 2-loop order}{79}
\contentsline {section}{\numberline {7}Extracting the physical informations: Extrapolations}{79}
\contentsline {subsection}{\numberline {7.1}The problem}{79}
\contentsline {subsection}{\numberline {7.2}General remarks about extrapolations and the choice of variables}{80}
\contentsline {subsection}{\numberline {7.3}Expansion about an approximation}{83}
\contentsline {subsection}{\numberline {7.4}Variational method and perturbation expansion}{84}
\contentsline {subsection}{\numberline {7.5}Expansion about Flory's estimate}{85}
\contentsline {subsection}{\numberline {7.6}Results for self-avoiding membranes}{86}
\contentsline {section}{\numberline {8}Other critical exponents and boundaries}{87}
\contentsline {subsection}{\numberline {8.1}Correction to scaling exponent $\omega $}{87}
\contentsline {subsection}{\numberline {8.2}Contact exponents}{88}
\contentsline {subsection}{\numberline {8.3}Number of configurations: the exponent $\gamma $}{89}
\contentsline {subsection}{\numberline {8.4}Boundaries}{91}
\contentsline {section}{\numberline {9}The tricritical point}{92}
\contentsline {subsection}{\numberline {9.1}Introduction}{92}
\contentsline {subsection}{\numberline {9.2}Double $\varepsilon $-expansion}{93}
\contentsline {subsection}{\numberline {9.3}Results and discussion}{97}
\contentsline {section}{\numberline {10}Variants}{99}
\contentsline {subsection}{\numberline {10.1}Unbinding transition}{99}
\contentsline {subsection}{\numberline {10.2}Tubular phase}{102}
\contentsline {section}{\numberline {11}Dynamics}{103}
\contentsline {subsection}{\numberline {11.1}Langevin-dynamics, effective field theory}{103}
\contentsline {subsection}{\numberline {11.2}Locality of divergences}{106}
\contentsline {subsection}{\numberline {11.3}Renormalization}{107}
\contentsline {subsection}{\numberline {11.4}Inclusion of hydrodynamic interaction (Zimm Model)}{110}
\contentsline {section}{\numberline {12}Disorder and non-conserved forces}{113}
\contentsline {subsection}{\numberline {12.1}The model}{115}
\contentsline {subsection}{\numberline {12.2}Field theoretic treatment of the renormalization group equations}{116}
\contentsline {subsection}{\numberline {12.3}Fluctuation-dissipation theorem and Fokker-Planck equation}{117}
\contentsline {subsection}{\numberline {12.4}Divergences associated with local operators}{118}
\contentsline {subsection}{\numberline {12.5}Renormalization of disorder (divergences associated with bilocal operators)}{120}
\contentsline {subsection}{\numberline {12.6}The residues}{122}
\contentsline {subsection}{\numberline {12.7}Results and discussion}{124}
\contentsline {subsection}{\numberline {12.8}Long-range correlated disorder and crossover from short-range to long-range correlated disorder}{128}
\contentsline {section}{\numberline {13}$N$-colored membranes}{129}
\contentsline {subsection}{\numberline {13.1}The $O(N)$-model in the high-temperature expansion}{130}
\contentsline {subsection}{\numberline {13.2}Renormalization group for polymers}{132}
\contentsline {subsection}{\numberline {13.3}Generalization to $N$ colors}{139}
\contentsline {subsection}{\numberline {13.4}Generalization to membranes}{140}
\contentsline {subsection}{\numberline {13.5}The arbitrary factor $\lowercase {c} (D)$}{145}
\contentsline {subsection}{\numberline {13.6}The limit $N\to \infty $ and other approximations}{145}
\contentsline {subsection}{\numberline {13.7}Some more applications}{147}
\contentsline {section}{\numberline {14}Large orders}{151}
\contentsline {subsection}{\numberline {14.1}Large orders and instantons for the SAM model}{152}
\contentsline {subsection}{\numberline {14.2}The polymer case and physical interpretation of the instanton}{155}
\contentsline {subsection}{\numberline {14.3}Gaussian variational calculation}{158}
\contentsline {subsection}{\numberline {14.4}Discussion of the variational result}{160}
\contentsline {subsection}{\numberline {14.5}Beyond the variational approximation and ${1/d}$ corrections}{164}
\contentsline {section}{\numberline {15}Conclusions}{165}
\contentsline {section}{\numberline {A}Appendices}{166}
\contentsline {subsection}{\numberline {A.1}Normalizations}{166}
\contentsline {subsection}{\numberline {A.2}List of symbols and notations used in the main text}{167}
\contentsline {subsection}{\numberline {A.3}Longitudinal and transversal projectors}{168}
\contentsline {subsection}{\numberline {A.4}Derivation of the RG-equations}{169}
\contentsline {subsection}{\numberline {A.5}Reparametrization invariance}{171}
\contentsline {subsection}{\numberline {A.6}Useful formulas}{171}
\contentsline {subsection}{\numberline {A.7}Derivation of the Green function}{173}
\contentsline {section}{\numberline {E}Exercises with solutions}{174}
\contentsline {subsection}{\numberline {E.1}Example of the MOPE}{174}
\contentsline {subsection}{\numberline {E.2}Impurity-like interactions}{175}
\contentsline {subsection}{\numberline {E.3}Equation of motion}{175}
\contentsline {subsection}{\numberline {E.4}Tricritical point with modified 2-point interaction}{176}
\contentsline {subsection}{\numberline {E.5}Consequences of the equation of motion}{177}
\contentsline {subsection}{\numberline {E.6}Finiteness of observables within the renormalized model}{178}
\contentsline {section}{\numberline {R}References}{179}

\bigskip
\bigskip

{This article will be published in volume 19 of ``Domb, Lebowitz'', 
``Phase Transitions and Critical Phenomena'', (C) Academic Press, London.
More information can be obtained from Academic Press or from
the author via email: {\tt\small wiese@theo-phys.uni-essen.de} or internet: {\tt\small 
http://www.theo-phys.uni-essen.de/tp/u/wiese/pubwiese.html}.
}
\end{document}